\documentstyle[11pt,fleqn]{article}

\newcommand{\beq}{\begin{eqnarray}}
\newcommand{\eeq}{\end{eqnarray}}
\newcommand{\al}{\alpha}
\newcommand{\ep}{\epsilon}
\renewcommand{\l}{\lambda}
\newcommand{\be}{\beta}
\newcommand{\nn}{\nonumber}
\newcommand{\y}{{\bf y}} 
\renewcommand{\k}{{\bf k}}                       
\newcommand{\x}{{\bf x}}                       
\newcommand{\xs}{{\bf x}'}
\newcommand{\xss}{{\bf x}''}        
\newcommand{\R}{\mbox{$I\!\!R$}}                 
\newcommand{\Z}{\mbox{$Z\!\!\!Z$}}               

\begin{document}

\hfill{\sl preprint - UTF 381 \\ hep-th/9607178 }
\par
\bigskip
\par
\rm

\par
\bigskip
\begin{center}
\bf
\LARGE

Euclidean Thermal Green Functions of Photons in Generalized Euclidean 
Rindler Spaces for any  Feynman-like Gauge.

\end{center}
\par
\bigskip
\par
\rm
\normalsize

\begin{center}\Large

Valter Moretti \footnote{e-mail:
\sl moretti@science.unitn.it}

\end{center}

\begin{center}
\large
\smallskip

 Dipartimento di Fisica dell'Universit\`a di Trento
\\and

\smallskip

 Istituto Nazionale di Fisica Nucleare,\\
Gruppo Collegato di Trento,\\ I-38050 Povo (TN)
Italia

\end{center}\rm\normalsize

\par
\bigskip
\par
\hfill{\sl July 1996}
\par
\medskip
\par\rm

\par
\begin{description}
\item{Abstract: }
\it \\ 
{\small  The thermal Euclidean Green functions for Photons
propagating in the Rindler wedge are computed employing an  Euclidean
approach within any covariant Feynman-like gauge.
This is done by generalizing  a formula which holds in the Minkowskian case.
 The coincidence of the found $(\be=2\pi)$-Green functions 
and the corresponding Minkowskian vacuum Green functions is discussed in 
relation to the remaining static gauge ambiguity already found
 in previous works.
Further  generalizations to more complicated manifolds are discussed.
 Ward identities are verified in the general case.} 
 
\par
\end{description}
\par
\medskip
\rm

\smallskip
\noindent{\sl PACS numbers:\hspace{0.3cm} 04.62.+v; 11.10.Wx; 11.15.-q  }
\par
\bigskip
\rm

\section{Introduction}
In principle, from the thermal Euclidean  Rindler Green functions 
of  quantum fields (also with spin) and   
employing standard different imaginary time analytic continuations,
 one can get either  the  {\em thermal} 
Feynman propagators in the Rindler space 
(for example see \cite{fullingruijsenaars} for the scalar case
and \cite{photrind} for the vectorial case)
 or the {\em non-thermal}    
Feynman propagators around a cosmic string at least for $\be\leq 2\pi$
(for example see \cite{linet} where generalizations to the case
$\be>2\pi$ are also studied).   \\
In fact, the (time periodic) Euclidean Rindler metric reads
\beq
ds^2 =  g_{ab} dx^a dx^b = \rho^2 d\theta^2 + 
d\rho^2 + dy^2 + dz^2 \label{metric}
\eeq
where $\rho \geq 0$ and $0\leq\theta\leq \beta \equiv 0$.
We may interpret $\theta$ as the imaginary Euclidean Rindler time 
and thus we may 
come back  in the Lorentzian Rindler space performing  an analytic 
continuation such as
$\theta \rightarrow -i\tau$. 
As it is well-known, the (Lorentzian) Rindler metric is the metric seen
 by uniformly 
accelerated observers, corresponding to the trajectories with 
$\rho,y,z =$ constants, in the portion of Minkowski space called (right) 
Rindler 
wedge. The relation between Minkowski coordinates $t,x,y,z$ 
and Rindler coordinates $\tau,\rho,y,z$ reads $t= \rho\sinh \tau$, $x= \rho
  \cosh\tau$, $\rho>0$, $\tau\in\R$.
 The right Rindler wedge is then defined as $x>|t|$.\\
 The value of $\be$ defines the temperature $T=1/\beta$ of the field quantum
 state for such  observers. 
Indifferently, we can perform the continuation
$z \rightarrow -it$ obtaining the metric in the Lorentzian section of
a cosmic string spacetime with time $t$.
 The constant $\beta$ is now connected to the 
linear mass density of the string $\mu = (2\pi-\be)/8\pi G$. \\
For the value $\be_U=2\pi$, corresponding to the well-known Unruh 
temperature \cite{unruh} as well as the disappearance of the
conical singularity at $\rho=0$ from the Euclidean and 
the Lorentzian string manifold,
  Minkowski vacuum
Green functions and $(\be=2\pi)$-Rindler Green functions are expected to
coincide. This is  due to the (local) coincidence of the corresponding
quantum states.
 A vast literature exists on this topic (see \cite{unruh,tutti} 
and \cite{haag}
 for  a recent
review)
also related to the physics of accelerated detectors \cite{birrel},
\cite{altri}. 
In other contexts,
 the metric (\ref{metric})  is 
also interesting because is a well-tried 
 approximation of the (Euclidean-section) metric near a Schwarzschild
 black hole event horizon. The particular value $\be_U$ is then
 related to Hawking's temperature of the black hole \cite{birrel}.

Let us focus our attention on the Euclidean Green function for a photon field
propagating in a Rindler wedge which has the correct periodic behaviour 
due to the Lorentzian KMS condition \cite{KMS,fullingruijsenaars}.
By the canonical quantization, summing over images and thus continuing
 into the
imaginary time, 
 Moretti and Vanzo  obtained in \cite{mova,photrind} the 
following (thermal) photon {\em Schwinger function}
 in the {\em Feynman gauge},
$\al$ being
 defined by $2\rho\rho'\cosh\al = \rho^2+\rho'^2+ |\x-\xs|^2$: 
\beq
S_{\be\: \theta\theta'}(x,x')
&=&\frac{1}{4\pi\be\sinh\al}
\frac{\cos\left(\frac{2\pi}{\be}
(\theta-\theta')\right)\sinh\al+\sinh\left[\left(\frac{2\pi}
{\be}-1\right)\al\right]}{\cosh\left(\frac{2\pi}{\be}\al\right)-
\cos\left(\frac{2\pi}{\be}(\theta-\theta')\right)}
+ \frac{1}{4\pi \be'}\:,\label{1''''}\\
S_{\be\: \rho\rho'}(x,x')&= &\frac{1}{\rho\rho'}S_{\be\:\theta\theta'}(x,x)
\label{1'''}\:,\\
S_{\be\: \theta\rho'}(x,x')&=&
\frac{-1}{4\pi\be\rho'}\frac{\sin\left(\frac{2\pi}{\be}
(\theta-\theta')\right)}{\cosh\left(\frac{2\pi}{\be}\al\right)-
\cos\left(\frac{2\pi}{\be}(\theta-\theta')\right)}
\:, \label{1''}\\
S_{\be\: \rho \theta'}(x,x') &=& -\frac{\rho'}{\rho}
S_{\be\: \theta\rho'}(x,x')
\:, \label{1'}\\
S_{\be\: yy'}(x,x')&=& S_{\be\:zz'}(x,x')=S_{\be}(x,x')
\label{1}\:.
\eeq
where $\be'=\be$ (see discussion below).
In \cite{24} similar results have been
obtained in the Lorentzian background of a cosmic string. 
An equivalent expression  for the case $\be=\be'$ is 
\beq
S_{\be\:aa'}(x,x') = V_{\be\:aa'}(x,x')\:\: S_\be(x,x') \label{1bis}\:,
\eeq
where 
\beq
V_{\be\:\theta\theta'} & = & \rho\rho' V_{\be\:\rho\rho'} 
= \rho\rho' \cosh \al \nonumber\:,\\
V_{\be\:\theta\rho'} & = & - \rho \rho'^{-1} V_{\be\:\rho \theta'} = 
- \rho \sin(\frac{2\pi}{\be}
(\theta-\theta')) \: \frac{\sinh \al }{ \sinh \frac{2\pi\al}{\be} }
\nonumber\\
 V_{\be\:yy'}
 & = & V_{\be\:zz'} =1 \:,\nonumber
\eeq
 all the remaining bi-vectors vanish. Here and above the  
function:
\beq
S_{\be}(x,x') =
\frac{1}{4\pi \be \rho\rho'\sinh \al}\left[ \frac{\sinh \left(\frac{2\pi}{\be}
\al\right)}
{\cosh\left(\frac{2\pi}{\be}\al\right) -\cos\left(\frac{2\pi}{\be}
(\theta-\theta')\right)}\right]
\label{2}
\eeq
is a well-known 
Rindler thermal Schwinger function for a massless scalar field
obtained by rotating into the imaginary time the 
Feynman propagator. This propagator is obtained by  the 
 sum over images after the canonical
quantization. 
These functions are Euclidean {\em Green functions}
 in the sense that they  satisfy
 the respective Green identities:
\beq
\nabla^{a} \nabla_{a} S_\be(x,x')
&=& -g(x)^{-1/2} \:\delta(x,x')\:, \label{x}\\
\nabla^{a} \nabla_{a} S_{\be\;bb'}(x,x')
&=& -g_{bb'}(x)\: g(x)^{-1/2}
\:\delta(x,x')
\:.\label{3}
\eeq
In the limit $ \be$ ( and $\be'$) 
$\rightarrow +\infty$ one recovers the non-thermal 
Schwinger functions calculated by the canonical 
quantization \cite{photrind}. 
The parameter $\be'$ which
 appears in $S_{\be\:\theta\theta'}$ and $S_{\be\:\rho\rho'}$
can be chosen different from the time period $\beta$ with no effect
for every physical quantity. This parameter gives raise
to  an ambiguity in defining all the Green functions\footnote{This is not a 
simple ambiguity due to an unfixed 
added constant to the Green functions because the term containing 
$\be'$ is a function
of $\rho$ and $\rho'$ in $S_{\be\:\rho\rho'}$.}. As discussed in 
\cite{mova,photrind} this is just a remaining static gauge ambiguity.
 Moreover, due to this gauge ambiguity 
 we have not the expected coincidence of Minkowskian 
and $(\be=2\pi)$-Rindler Green 
function {\em also} for $\beta'=\beta (=2\pi)$. One must take $\beta'= +\infty$
in order to reach such a coincidence. 
However, the physics seems to be safe because
 this coincidence is restored, non depending on the value of $\be'$,
for the Lorentzian two-point function  as well as the Feynman propagator
when they  work on Lorentz and 
physical photon wavefunctions \cite{mova,photrind}. 
The presence of several problems in studying photon theory in a Rindler
wedge (also related with the gauge invariance) has been encountered 
in other contexts \cite{kabat,ielmo}. 
As far as the Green functions are concerned, 
 it is interesting to  evaluate the photon Green functions 
 employing methods different from the sum over images and 
check if a different value for $\be'$ is selected
(for instance the more natural value $\be'=+\infty$). 
Furthermore, it is also intersting to  check  whether or not
 the previous
problems are specific features of the Feynman gauge only. 

The  result of the present paper,  which generalizes a well-known formula 
holding in the Minkowskian case, is a  
development of the previous papers  because the Green function 
expression we shall get holds
for any (Feynman-like) covariant gauge  and 
for a generalized Euclidean Rindler space 
${\cal M}\times \R^2$ endowed with the natural product 
metric. 
 ${\cal M}$ is a generally curved two-dimensional Euclidean 
 manifold which, in the Rindler case, reduces 
 to  ${\cal C}_\beta$, i.e. a cone of an angular deficit $2\pi-\beta$.
However, differently from \cite{mova} and \cite{photrind}, our result will
 arise from 
a direct  Euclidean approach instead of the canonical quantization. 
For this reason we prefer the term Euclidean Green function instead of 
Schwinger function.\\
We shall see that the Green functions we found in the Rindler wedge
contain the same static gauge ambiguity previously
discussed only for the Feynman gauge 
  and we shall get the coincidence of $(\beta = 2\pi)$-Euclidean Rindler 
Green functions and Euclidean Minkowski Green functions
except for the usual discrepancy corresponding to the presence of
 unphysical photons in the Lorentzian theory.
Finally, we shall prove the Ward identities for the general case.

\section{Euclidean Approach for Feynman Gauge}

Let us start by considering an Euclidean Rindler wedge of time period $\beta$:
${\cal C}_\be\times \R^2$. 
From now on Latin indices as $a, b, c, d$ are for the whole manifold,
 Greek indices are for the pure cone and Latin indices as $i,j...= y,z$
are for the remaining $\R^2$. The Euclidean Maxwell equations, 
in a generic Feynman-like covariant gauge parametrized
by $\eta \in \R$, read
\beq
\left[ \delta^{b}_{a}\Delta + R_a^{b} -
(1-\frac{1}{\eta})\nabla_{a}\nabla^{b} \right] A_b = 0 \label{4}\:.
\eeq
$\Delta = \nabla_c\nabla^c$
 is the Laplace-Beltrami vectorial operator built up using 
covariant derivatives. Obviously, when $\eta=1$ 
one recovers the usual Feynman gauge Maxwell 
equations.\\
In the case of the Rindler space  $R_a^b=0$ holds
except for $\rho=0$, however we shall employ normal modes which vanish
in those points and thus we shall omit $R_a^b$. Later,  we shall consider
also  that term in more complicated manifolds  employing 
the Hodge-de Rham formalism.\\
We want to get a Green function of the previous operator by the usual
sum of normal modes products. Let us first consider the case $\eta=1$
and prove that this leads us to the Schwinger function in Eq.(\ref{1bis}).
An useful complete  eigenfunctions set
of the vectorial Laplace-Beltrami operator (corresponding to $\eta=1$ 
and $R^b_a=0$ in Eq.(\ref{4}))  is that discussed by 
Kabat \cite{kabat}:
\begin{eqnarray}
A_a^{(I,n\lambda{\bf k})}&:=&(0, 0, \phi, 0)\label{5'},\\
A_a^{(II,n\lambda{\bf k})}&:=&(0, 0, 0, \phi)\label{5''},\\
A_a^{(III,n\lambda{\bf k})}
&:=&\frac{\sqrt{g}}{\lambda}
\epsilon_{\mu\nu}\partial^\nu\phi=
\frac{1}{\lambda}
(\rho\partial_\rho\phi,-\frac{1}{\rho}\partial_\theta\phi,0,0),\label{5'''}\\
A_a^{(IV,n\lambda{\bf k})}&:=& \frac{1}{\lambda}
 \partial_\mu \phi=
\frac{1}{\lambda}
(\partial_\theta\phi,\partial_\rho\phi,0,0)\: \label{five}
\end{eqnarray} 
 $\sqrt{g}\epsilon_{\mu\nu}$ is the Levi-Civita pseudo-tensor on the
cone and $\phi=\phi^{(n\lambda{\bf k})}(x)$ defines the complete
scalar eigenfunctions set of a  self-adjoint
extension of the scalar Laplacian on $C_\beta\times \R^2$
 with eigenvalues $-(\l^2+\k^2)$ \cite{kay,kabat}:
\begin{eqnarray} 
\phi^{(n\lambda{\bf k})}(x)&:=&\frac{1}{2\pi}\sqrt{\frac{\l}{\be}}
e^{i\k \x}
e^{i\frac{2\pi n}{\beta}\theta} J_{\nu_n}(\lambda \rho),\hspace{5mm}
n\in\Z \:; \,\,\lambda\in \R^+;\,\, {\bf k}=(k_y,k_z) \in \R^2
\label{6'}\:.
\end{eqnarray}
Here $J_{\nu_n}$ is the Bessel function of first kind and
$\nu_n=\frac{2\pi|n|}{\beta}$.  The  previous modes are normalized according
to
\begin{eqnarray}
 \int d^4x\,
\sqrt{g} \,g^{ab}
A_a^{(m',n'\lambda'{\bf k}')\ast}A_b^{(m,n\lambda{\bf k})}
&=&\delta_{m'm} \delta_{n'n}\delta^{(2)}({\bf k}-{\bf k}')
\delta(\lambda-\lambda')\:,\label{norm}\\
 \int d^4x\,
\sqrt{g} 
\phi^{(n'\lambda'{\bf k}')\ast}\phi^{(n\lambda{\bf k})}
&=&\delta_{n'n}
\delta^{(2)}({\bf k}-{\bf k}')
\delta(\lambda-\lambda')\:, \label{norm'}
\end{eqnarray}
In the considered case $\eta=1$, all the previous  eigenfunctions of the
vectorial Laplacian  
correspond to the eigenvalues 
$-(\lambda^2+ \k^2)$. 
A Green function of this vectorial operator, hence satisfying Eq.(\ref{3}),
can be  built up by the found modes as:
\beq
G_{\be\:aa'}(x,x') 
&:=&  \int_0^{+\infty} d\lambda\:
\int_{\R^2}d\k \sum_{n\in\Z} \sum_{m= I}^{IV}
\frac{A_a^{(m,n\lambda{\bf k})}(x) 
A_a^{(m,n\lambda{\bf k})\ast}(x')}{\lambda^2+
 \k^2} \label{7}\:;
\eeq	
we can equivalently write
\beq
G_{\be\:aa'}(x,x') &=&  \int_0^{+\infty} \frac{d\lambda}{\l^2}\:
\int_{\R^2}d\k \sum_{n\in\Z}  D_{aa'}
\frac{\phi^{(n\lambda{\bf k})}(x)
\phi^{(n\lambda{\bf k})\ast}(x')}{\lambda^2+
 \k^2}\label{7'}\:,
\eeq
where  we used the definitions: 
\beq 
D_{\theta\theta'} &=& \rho\rho' D_{\rho\rho'} =
 \partial_\theta\partial_{\theta'} + 
\rho\rho' \partial_\rho \partial_{\rho'}\:;\nonumber\\
 D_{\theta\rho'} &=&-\rho \rho'^{-1}
D_{\rho \theta'}=\rho'^{-1}[\partial_\theta \rho'\partial_{\rho'}- 
\partial_{\theta'}
\rho\partial_\rho]\:; \nonumber \\ 
D_{yy'}&=&D_{zz'}=\l^2\:.\nonumber
\eeq
 All the remaining components vanish.
By employing the following regularization:
\beq
\lambda^{-2} (\l^2+\k^2)^{-2} = \l^{-2}(\k^2+\ep^2)^{-2}
- (\k^2+\ep^2)^{-2}(\l^2+\k^2)^{-2}\:\:\:\: \mbox{ as}\:\:\:\: 
\ep \rightarrow 0 \nonumber\:,
\eeq 
one can obtain as far as the conical components are concerned:
\beq
G_{\be\:\mu\mu'}(x,x')&=& -\int_0^{+\infty} d\lambda\:
\int_{\R^2}d\k \sum_{n\in\Z}  \frac{D_{\mu\mu'}}{\k^2+\ep^2}
\frac{\phi^{(n\lambda{\bf k})}(x)\phi^{(n\lambda{\bf k})\ast}(x')}{\lambda^2+
 \k^2}+\nn\\
& & +\int_0^{+\infty} d\lambda\:
\int_{\R^2}d\k \sum_{n\in\Z}  \frac{D_{\mu\mu'}}{\k^2+\ep^2}
\frac{\phi^{(n\lambda{\bf k})}(x)\phi^{(n\lambda{\bf k})\ast}(x')}{\lambda^2}
\:.
\label{8}
\eeq
The second term in the right hand side of Eq.(\ref{8})
can be written:
\beq
\left(\frac{1}{4\pi^2} \int_{\R^2}d\k \frac{e^{\k (\x-\xs)}}{\k^2+\ep^2}
\right) D_{\mu\mu'} \int_0^{+\infty} d\lambda\:\sum_{n\in\Z} 
\frac{\varphi^{(n\lambda)}(\theta,\rho)\varphi^{(n\lambda)\ast}(\theta'\rho')}{
\lambda^2}\label{9}\:,
\eeq
where \beq
\varphi^{(n\lambda)}(\theta,\rho) = \sqrt{\frac{\l}{\be}}
e^{i\frac{2\pi n}{\beta}\theta} J_{\nu_n}(\lambda \rho)
\eeq 
is an eigenfunction
of the scalar Laplacian on the pure cone ${\cal C}_\be$
 with eigenvalue $-\l^2$.
Hence,  the remainder after $D_{\mu\mu'}$ in Eq.(\ref{9}) coincides with 
 a Green function of the Laplacian on the pure cone:  
\beq
G^c_\beta(x,x')=
-(2\be)^{-1}
\ln|\rho^2+\rho'^2-2\rho\rho'\cos(\theta-\theta')|
\:.\nonumber
\eeq
 Furthermore, by a direct check on the right hand side of the previous
expression,
 one  
finds:
\beq
D_{\mu\mu'} G^c_\beta(x,x')=0\:,    \nonumber
\eeq
 and thus
only the first term in the right hand side of Eq.(\ref{8}) survives.
Then we have:
\beq
G_{\be\:\mu\mu'}(x,x')= -\int_0^{+\infty} d\lambda\:
\int_{\R^2}d\k \sum_{n\in\Z}  \frac{D_{\mu\mu '}}{\k^2+\ep^2}
\frac{\phi^{(n\lambda{\bf k})}(x)\phi^{(n\lambda{\bf k})\ast}(x')}{\lambda^2+
 \k^2} \label{10}\:.
\eeq 
As we expect from the general theory, one can  prove the further identity
\beq
G_\be(x,x') :=  \int_0^{+\infty} d\lambda\:
\int_{\R^2}d\k \sum_{n\in\Z} 
\frac{\phi^{(n\lambda{\bf k})}(x)\phi^{(n\lambda{\bf k})\ast}(x')}{\lambda^2+
 \k^2} \equiv S_{\beta}(x,x')\:, \label{11}
\eeq
where the latter right hand side is just the scalar
 Schwinger function defined in Eq.(\ref{2}), obtained from the canonical 
quantization and the sum over images. This proves, through Eq.(\ref{7'}), that 
$G_{\be\:ij}(x,x') = S_{\be\:ij}(x,x')$. We can employ Eq.(\ref{11})
to further develop the expression in Eq.(\ref{10}).  
Following the same way used in \cite{photrind} for the non-thermal case we can
write Eq.(\ref{10}) as
\beq
G_{\be\:\mu\mu'}(x,x')= -\frac{1}{2\pi} \int_{\R^2} d\xss \ln 
\frac{|\xss|}{\mu_0}D_{\mu\mu'} S_{\be}(\theta-\theta',\rho,\rho',\xss
- (\x-\xs))
\label{12}\:,
\eeq
where $\mu_0$ is an unimportant constant defined in \cite{photrind}.
Moreover we can prove the following identities: 
\beq
D_{\theta\theta'}S_\be &=& \rho\rho' D_{\rho\rho'} S_\be =
 \rho\rho' \nabla^2_{\x}[\cosh 
\al \: S_\be(\theta,\theta',\rho,\rho',\x) ] \nonumber\:, \\
 D_{\theta\rho'} S_\be &=& - \rho'\rho^{-1}D_{\rho \theta'}S_\be
 = -\rho \sin(\frac{2\pi}{\be}
(\theta-\theta'))\: \nabla^2_{\x}[\sinh \al\: 
S_\be(\theta,\theta',\rho,\rho',\x)/ \sinh \frac{2\pi\al}{\be}
] \nonumber\:.
\eeq
 Using these in Eq.(\ref{12}) and reminding also
that for smooth $\R^2$ functions ($c\in \R^+$ fixed):
\beq
\int_{\R^2} d{\bf x} \ln (|\x|/c) \nabla^2_{\x} g(\y- \x) = -2\pi
g(\y) \nonumber\:,
\eeq
 one can prove that 
\beq
G_{\be\:\mu\mu'}(x,x')
= S_{\be\:\mu\mu'}(x,x') 
\nonumber
\eeq
(where we supposed $\be'=\be$) by a direct 
comparison with the expression of $S_{\be\:\mu\nu}(x,x')$ given in
Eq.(\ref{1bis}).
Summarizing, we have finally proved that 
\beq
G_{\be\:aa'}(x,x') = S_{\be\:aa'}(x,x') \label{14}
\eeq
Thus, the same Green function calculated through the canonical quantization and
the sum over images re-arises, with the same value $\be$ for $\be'$.
 The non-thermal Schwinger function obtained by the canonical 
quantization arises as 
 the limit as $\be\rightarrow +\infty$ of $G_{\be\:aa'}$.
 
\section{Euclidean Approach for any Feynman-like Covariant Gauge}

Let us now consider the case $\eta \neq 1$ in the operator in the 
left hand side 
of  Eq.(\ref{4}). 
An useful set of eigenfunctions
 normalized as in Eq.(\ref{norm}) and
 built up by linear combinations of the previous Kabat's modes  reads
 \cite{ielmo}:
\begin{eqnarray}
A_a^{(I,n\lambda{\bf k})}&=&\frac{1}{k}\epsilon_{ij}\partial^j\phi=
\frac{1}{k}(0,0, ik_z\phi,-ik_y\phi),\label{15'}
\\
A_a^{(II,n\lambda{\bf k})}&=&\frac{\sqrt{g}}{\lambda}\epsilon_{\mu\nu}
\partial^\nu\phi=
\frac{1}{\lambda}(\rho\partial_\rho\phi,-\frac{1}{\rho}
\partial_\theta\phi,0,0),
\label{15''}
\\
A_a^{(III,n\lambda{\bf k})}&=&\frac{1}{\sqrt{\lambda^2+{\bf k}^2}}
(\frac{k}{\lambda}\partial_\mu-\frac{\lambda}{k}\partial_i)\phi=
\frac{1}{\sqrt{\lambda^2+{\bf k}^2}}
(\frac{k}{\lambda}\partial_\theta\phi,\frac{k}{\lambda}\partial_\rho\phi,
-\frac{\lambda}{k}\partial_y\phi,-\frac{\lambda}{k}\partial_z\phi),
\label{15'''}
\\
A_a^{(IV,n\lambda{\bf k})}&=&\frac{1}{\sqrt{\lambda^2+{\bf k}^2}}
\partial_a\phi=\frac{1}{\sqrt{\lambda^2+{\bf k}^2}}
(\partial_\theta\phi,\partial_\rho\phi,\partial_y\phi,\partial_z\phi),
\label{15}
\end{eqnarray}
Above, $\phi := \phi^{(nk\l)}(x)$ previously defined, $k:= |\k|$ 
and $\epsilon_{ij}$ is 
 the Levi-Civita pseudo-tensor on $\R^2$ in Cartesian coordinates.
The first three eigenfunctions  satisfy $\nabla^a A_a=0$
and have eigenvalue 
$\varepsilon^{(\eta)}_{I} = \varepsilon^{(\eta)}_{II} = 
\varepsilon^{(\eta)}_{III} =
-(\lambda^2+{\bf k}^2)$, while $A_a^{(IV)}$ 
 has eigenvalue  $\varepsilon^{(\eta)}_{IV}= 
-(\lambda^2+{\bf k}^2)/\eta$.\\
Few calculations, involving only Eq.(\ref{7}) and the form of the employed 
modes in terms of
 the scalar modes, lead  quite easily us to\footnote{Primed index
derivatives act on primed arguments.}:
\beq
G^{(\eta)}_{\be\:aa'}(x,x')
&:=&  \int_0^{+\infty} d\lambda\:
\int_{\R^2}d\k \sum_{n\in\Z} \sum_{m= I}^{IV}
\frac{A_a^{(m,n\lambda{\bf k})}(x)
A_a^{(m,n\lambda{\bf k})\ast}(x')}{ \varepsilon^{(\eta)}_{y}} =\label{S} \\
& = & G_{\be aa'}(x,x') + (\eta -1) \partial_a \partial_{a'}
\int d\lambda\:
\int d\k \sum_{n\in\Z}
\frac{\phi^{(n\lambda{\bf k})}(x)\phi^{(n\lambda{\bf k})\ast}(x')}{(\lambda^2+
 \k^2)^2}\:. \label{S'}
\eeq
Hence, due to Eq.s (\ref{11}) and (\ref{norm'}) we have our main result:
\beq
 G^{(\eta)}_{\be\:aa'}(x,x')
 = G_{\be aa'}(x,x') + (\eta -1) \partial_a \partial_{a'}
[G_\be * G_\be](x,x') \:,\label{16}
\eeq
or, equivalently:
\beq
 G^{(\eta)}_{\be\:aa'}(x,x')=
 S_{\be aa'}(x,x') + (\eta -1) \partial_a \partial_{a'}
[S_\be * S_\be](x,x')
\label{16'}\:.
\eeq 
Denoting by  ${\bf s}$  the three spatial variables,
 the previous convolution is defined as 
\beq
 [f*f'](x,x') := \int d^4y \sqrt{g(y)} \: f(x,y) f'(y,x') 
=  \int d^3{\bf s} \sqrt{g({\bf s})} \int_0^\beta d\theta 
 \: f(x,({\bf s},\theta)) f'(({\bf s},\theta), x') \nn \:,
\eeq
Despite of a different manifold structure and a different choice for the 
Euclidean time, we have found  the same structure of 
$\eta$-parametrized Green functions holding in the 
Euclidean section of Minkowski's space for the vacuum 
 $\eta$-parametrized Minkowskian Green functions.

\section{Discussion on a Remaining Gauge Ambiguity}

Few remarks on Eq.(\ref{16}). Let us prove that, in the  case $\eta \neq 1$, 
the same features of the case $\eta=1$ \cite{mova,photrind} re-appear.
$G_{\be\:aa'}(x,x')
(= S_{\be\:aa'}(x,x'))$ which appears in Eq.(\ref{16}) contains the two static 
terms
$1/4\pi\be' $ and $ 1/4\pi \rho\rho'\be$ 
 evaluated at $\be'=\be$,
 respectively in $S_{\be\:\theta\theta'}(x,x')$ and $S_{\be\:\rho\rho'}(x,x')$.
\\
In our Euclidean approach these {\em static} terms arise from the zero modes 
$\varphi^{(0 \l)}(\rho)$ of the Laplacian on the pure cone. 
In a different context, Kay and Studer pointed out other subtleties related
 to these zero  modes \cite{kay}.         \\
We can write such terms by a pure gauge term form as
\beq
\delta G_{\be'\:aa'}( = \delta S_{\be'\:aa'})
:= \partial_a\partial_{a'} \Phi(x,x')_{\be'=\be}\:, \nonumber 
\eeq
where 
\beq
\Phi(x,x')_{\be'} := (4\pi\be')^{-1}(\theta\theta'+ \ln\rho\ln\rho')\nonumber
\:.
\eeq
Thus, the Green function for the strength field
obtained from vectorial Green function employing the definition:
\beq
F_{ab} := \nabla_aA_b -\nabla_bA_a = \partial_aA_b -\partial_bA_a\nonumber
\eeq
and the corresponding physical quantities as the stress tensor already
 calculated in \cite{mova}, do
 not depend on the value of $\be'$ as well as  on the value of $\eta$.
\\
Furthermore, we may quite simply obtain\footnote{Remind that 
covariant derivatives
commute due to the flatness of the manifold.}
 \beq \nabla_a\nabla^a \Phi(x,x')_{\be'}=0    \:\:\:\:
\mbox{and}\:\:\:\: \nabla_b\nabla^b \delta G_{\be'\:aa'}=0 \nonumber\:.
\eeq
Due to these properties, non depending on the values of
 $\be' (\neq \be$ in general), $\delta G_{\be'\:aa'}$ gives no contribution
in verifying the $\eta-$Green equation: 
\beq
[\delta^b_a\nabla^{c} \nabla_{c} - (1-\frac{1}{\eta}) \nabla_a\nabla^b]
G_{\be\;bb'}(x,x')
= -g_{ab'}(x)\: g(x)^{-1/2}
\:\delta(x,x') \label{green3}\:.
\eeq
 Similarly it gives no contribution in verifying
 the $(\eta=1)$-Ward identity  necessary for the  BRST 
invariance\footnote{In this context the scalar Green function $G_\be$
has to be interpreted as the ghosts Green function.} \cite{mova,photrind}  
\beq
\nabla^a G_{\be\:aa'}(x,x') + \nabla_{a'}G_\be(x,x') = 0 \:. \label{ward1}
\eeq 
The above identity can be proved using directly
the definitions in Eq.s (\ref{7}), (\ref{11}) and
Kabat's modes (\ref{5'})-(\ref{five}) (without expliciting the particular 
expression of the scalar eigenfunctions).
Starting from Eq.(\ref{ward1}),  non depending on the value of
$\be'$, due to Eq.(\ref{x}) and Eq.(\ref{16}), 
 the $\eta-$Ward identity (also necessary  for the  BRST invariance):
\beq
\nabla^a G^{(\eta)}_{\be\:aa'}(x,x') + \eta 
\nabla_{a'}G_\be(x,x') = 0  \label{ward2}
\eeq 
can be proved.\\
Hence, as in the case $\eta=1$, the term $\delta G_{\be'\:aa'}$ represents
a remaining static gauge ambiguity 
which does not affect the physics non depending
on the value of $\be'$. \\
Few words on the mathematical structure of the found Green functions.\\
As it was
discussed in \cite{photrind}, if $\be\neq 2\pi$ no value of $\be'$ produces 
an everywhere defined Green function  due to the 
conical singularity at $\rho$ (or $\rho'$) $=0$,
 thus no value of $\be'$ is preferred in a mathematical 
context. Conversely, when $\be=2\pi$, the choice $\be'=\be$, 
and {\em only} that, produces
a Green function well-definable on the whole manifold.\\

\section{Vacua Coincidence at $\be=2\pi$}

Let us address ourselves to what happens 
to the previous Green functions at $\be= 2\pi$. Remind that, as found in 
\cite{photrind} 
\beq
S_{2\pi\:aa'}(x,x') =  
S^{\scriptsize \mbox{Minkowski}}_{\infty\:aa'}(x,x') + 
\delta S_{2\pi\:aa'}(x,x')      \:. \nonumber
\eeq
 where the previous 
 Minkowski Schwinger function is obtained by the canonical quantization 
and is referred to the Minkowski vacuum. 
Taking into account that conversely 
\beq
S_{2\pi}(x,x')  =
S^{\scriptsize \mbox{Minkowski }}_{\infty}(x,x')
\nonumber
\eeq
 and reminding 
the well-known form
of the Euclidean  Minkowski vacuum Green function in any covariant gauge,  
we obtain:
\beq
G^{(\eta)}_{2\pi\:aa'}(x,x') = 
S^{(\eta)\:\:\scriptsize \mbox{Minkowski}}_{\infty\:aa'}(x,x') +
\delta G_{2\pi\:aa'}(x,x') \:. \nn
\eeq
Notice that
\beq
  \delta G_{\be'\:aa'}
= \partial_a\partial_{a'} \Phi(x,x')_{\be'}\:,\nonumber
\eeq
 as well as the convolution 
term 
in Eq.(\ref{16}),
 gives a vanishing result due to the divergence theorem, 
working as an integral kernel on compact support 
  smooth test functions $J^a(x)$ satisfying 
$\nabla_a J^a(x)=0$. 
 Thus, the broken coincidence of Green functions
is restored in the sense of distributions working on vanishing divergence 
test functions. The Lorentzian  analogue of this fact is the distributional 
coincidence of the Lorentzian Green functions working on
  three-smeared wavefunctions of  Lorentz and  physical photons
within a Gupta-Bleuler-like formalism  (see
 \cite{mova} and \cite{photrind} and also Appendix C therein). \\

\section{Generalizations in more General Manifolds}

Generalizations to manifolds as ${\cal M}\times \R^2$ endowed
with the natural product metric are straightforward. We can consider 
${\cal M}$ as a generally Euclidean curved two-manifold and define 
coordinates $x^\mu \equiv (\theta, \rho)$  on that. 
We could also suppose that $\partial_\theta$
defines a Killing vector with compact orbits of period $\beta$ to
be interpreted as the inverse of a temperature; however such a supposition 
is not so strictly necessary. For this generalized case
(as well as in completely general manifolds), Euclidean Maxwell
equations read
\begin{eqnarray}
\left[ -\Delta_1 + (1- \frac{1}{\eta}) d_0 \delta_0
\right] A = 0 
\label{hodge}\:,
\end{eqnarray}
where $\Delta_1 = d_0\delta_0 +\delta_1 d_1$ is the  Hodge Laplacian
for 1-forms ($\delta_n := d_n^\dagger $ w.r.t. the Hodge scalar
product.).
The eigenfunctions of the operator appearing in the
above equation can  be still written as in the former right hand side of 
Eq.s(\ref{15'})-(\ref{15}), provided 
\beq
\phi= \frac{1}{2\pi} e^{i\k\x} {\bf J}_{n,\lambda}(x^\mu)
\nonumber\:,
\eeq
 where
${\bf J}_{n,\lambda}(x^\mu)$ is an eigenfunction of 
 the 0-forms Hodge Laplacian
$\Delta^{\cal M}_0$ on ${\cal M}$, with eigenvalue $\lambda^2$. Employing
a bit of n-forms algebra, one can obtain in our manifold the same
eigenvalues found in ${\cal C}_\beta \times \mbox{R}^2$.
Furthermore, once again $\delta_0 A^{(m)} = 0$,  namely $ \nabla^a
A_a^{(m)}= 0 $, in the 
cases $m = I, II, III$.   \\
Starting from  the definitions in 
Eq.(\ref{11})  and Eq.(\ref{S}) one can simply prove Eq.s (\ref{S'}) and
(\ref{16})
using the considered modes without expliciting the
particular form of the scalar modes ${\bf J}_{n,\lambda}(x^\mu)$.
Similarly, one can prove the Ward identities in Eq.(\ref{ward2}) directly
by Eq.(\ref{S}). 
Notice that  particularly trivial cases are ${\cal M} = \R^2$ and
 ${\cal M} = S^1\times\R$.
In these cases we trivially find respectively the expressions of the
vacuum  Minkowski Schwinger function and the thermal Minkowski Schwinger 
function in terms of the corresponding scalar Schwinger function and the
Feynman Gauge Schwinger function.\\
Just a remark for the case of a completely general manifold
generally different from ${\cal M}\times \R^2$. Suppose that
in a general manifold
$G^{(\eta=1)}_{aa'}(x,x')$ and $G^{\scriptsize \mbox{scalar}}(x,x')$ 
are Green functions of $\Delta_1$
and $\Delta_0$ respectively. One can simply prove, employing the n-forms
theory, that $G^{(\eta)}_{aa'}$ given in Eq.(\ref{16})
(where $G_{\be\:aa'} \rightarrow G^{(\eta=1)}_{aa'}$ and $G_\be \rightarrow 
G^{\scriptsize \mbox{scalar}}$)
  both can still  be a Green function of the operator in
 Eq.(\ref{hodge}) and  can still satisfy the $\eta-$Ward identity of
 Eq.(\ref{ward2}). In fact, this holds 
if, and only if, the former two Green functions  satisfy the $\eta=1$ Ward
identity of Eq.(\ref{ward1}) which now reads 
\beq
\delta_0 G^{(\eta=1)}(x,x') -
d'_1 G^{\scriptsize \mbox{scalar}}(x,x') =0  \nonumber\:.
\eeq

\par \section*{Acknowledgments}
I would like to thank  Luciano Vanzo and Devis Iellici 
 for some discussions on the topics of this paper.


\end{document}